\documentclass[aps,prl,twocolumn]{revtex4}

\usepackage[dvips]{graphicx}
\usepackage[]{amsmath}
\usepackage[]{amssymb}
\usepackage[]{amsfonts}
\usepackage[]{latexsym}

\usepackage[]{float}

\newcommand{\mb}[1]{\mbox{\boldmath $#1$}}

\newcommand{\eas}[0]{\begin{eqnarray*}}
\newcommand{\eae}[0]{\end{eqnarray*}}
\newcommand{\les}[0]{\begin{equation}}
\newcommand{\lee}[0]{\end{equation}}
\newcommand{\leas}[0]{\begin{eqnarray}}
\newcommand{\leae}[0]{\end{eqnarray}}

\begin{document}


\title{
A Characterization of Topological Insulators:\\
 Chern Numbers for a Ground State Multiplet
} 


\author{Y. Hatsugai}
\email[]{hatsugai@pothos.t.u-tokyo.ac.jp}
\affiliation{Department of Applied Physics, University of Tokyo,\\
7-3-1, Hongo, Bunkyo-ku, Tokyo 113-8656, JAPAN}


\date{ Nov. 18, 2004, Nov. 22}

\begin{abstract}
We propose to use generic Chern numbers for a characterization 
of topological insulators.
It is suitable for a numerical characterization of
low dimensional quantum liquids 
where strong quantum fluctuations
prevent from developing conventional orders. 
By twisting parameters of boundary conditions, 
the non-Abelian Chern number are defined for a few low lying states
near the ground state in a finite system, which is a ground state
multiplet with a possible (topological)
 degeneracy.
We define the system as a topological insulator
when energies of the multiplet are well separated from the above.
Translational invariant twists up to a unitary equivalence
are crutial to pick up only bulk properties without edge states.
As a simple example, the setup is applied for a
two-dimensional $XXZ$-spin system with an ising anisotropy where
the ground state multiplet is composed of doubly almost degenerate states.
It gives a vanishing Chern number due to a symmetry. 
Also Chern numbers for the generic fractional quantum Hall states are 
discussed shortly.
\end{abstract}


\maketitle

%


A crucial role of phases is one of  intrinsic features of  quantum mechanics
 and has a long history of investigation.
Among them,
 those which have intrinsic geometric origins
are now understood as 
geometrical phases\cite{Shapere89}.
Aharonov-Bohm effects and  Dirac monopoles
are typical and classic examples where 
the geometrical phases are fundamental.
Geometrical features of gauge theories
are another prototype\cite{eguchi80}.
Also a discovery of Berry's phases reveals 
that  the geometric phases and the gauge structure
are closely related and  derived 
by 
restricting a physical Hilbert space\cite{berry84,Wilczek84}.

On the other hand, with an idea of order parameters,
 symmetry breaking 
is one of the most fundamental concepts  
in modern physics. 
Quite successfully,
 this standard setup can characterize most of ordered states
and describe phase transitions and critical phenomena.
However in low dimensional quantum systems, 
such as electrons 
 with strong correlation and  spins, 
 quantum fluctuations prevent from developing conventional
orders even at a zero temperature. 
In these systems, quantum phases of manybody ground states vary wildly
in space and time,
which destroy the standard orders. 

Of course, the wild quantum phases  are not random but obey 
some hidden restriction rules and reflect 
features of the quantum mechanical wave function.
Some of them are well known today such as Marshall sign rules
in the spin systems\cite{Marshall55} and fractional statistics 
of quasi-particle (hole) wavefunctions in the 
fractional quantum Hall effect (FQH)\cite{Laughlin83}.
String order parameters in the Haldane spin 
chains\cite{denNijs89,Girvin89,Hatsugai92}
and the quasi off-diagonal order in the FQH\cite{Girvin87}
are also discussed based on the feature.
Generically quantum states with the characteristic 
geometric phases are considered to 
possess non-trivial topological orders\cite{Wen89}.

Recently wide variety of interesting and physically important phenomena
have been 
understood based on a concept of the topological order.
Some of them include quantum Hall effects\cite{Laughlin81, Thouless82},
solitons in  polyacetylens\cite{ssh}, adiabatic transports of
charge and spins\cite{pump-th},
itinerant magnetism and spintronics\cite{tokura-chiral} and
 anomalous Hall conductances\cite{aHallMac,spintronics,Haldane04},
polarizations in insulators\cite{ksv},
 the two dimensional carbon sheets\cite{Haldane88honey,sr-yh},
anisotropic superconductors\cite{smf,sr-yh,Hatsugai04a},
and string-net condensations\cite{WenString}.
They are under active studies.

A local phase of the manybody wave function is arbitrary but
there is some correlation with the phases of its neighbors,
which  brings some gauge structures.
In these view points,
  templates of such systems are 
 quantum Hall states, especially 
integer quantum Hall (IQH) states.
There are apparently different QH states with different quantized
Hall conductances. 
However, any symmetries are not broken among the states but they are clearly
different physical systems. 
These states are characterized by the quantized Hall conductances
which have an intrinsic topological origin\cite{Laughlin81}.
The topological origin of the Hall conductance is clear by 
the Chern number expression\cite{Thouless82,Kohmoto85,ntw}.
Based on the observation, 
we {\em propose to characterize the topological orders by the
generic Chern numbers}\cite{Hatsugai04}.  
The Chern numbers in topological ordered systems are kinds of 
order parameters in conventional ordered phases.
In the same manner as
 the usual phase transition is characterized by a sudden change of 
order parameters, topological phase transitions are 
characterized by a discontinuous change of the Chern numbers.

To have a well defined Chern number, we need an existence of a
generic gap\cite{Hatsugai04}.  
{\em Topological insulators } are defined as physical systems
with this generic energy gap.
Then the  Chern numbers are always integers and the topological 
phase transition is characterized by a discontinuous change 
of the Chern numbers
which are always integers. 
This integral property of the Chern number implies 
a stability of the characterization against a small perturbation.
However,
as in the case of the edge states of the quantum Hall effect\cite{Halperin82} and
Kennedy's triplet states in the Haldane spin chains\cite{Kennedy90},
 the topological ordered state are quite sensitive to a
geometrical change of the physical system such as
an  existence of edges and boundaries. 
It contrasts with the conventional order where boundary conditions 
are always negligible in the thermodynamic limit. 
Therefore a translational invariance is fundamentally 
important to describe topological ordered states.
On the other hand,  in many cases, 
as far as physical observables are concerned, 
a topological order is hidden in a bulk
and  only reveals its physical significance near boundaries of the
system.

Generically speaking, 
to define the Chern
numbers $C$ for a physical (many particle) wave function, $\psi$,
 we need to require the wave function to depend on multiple parameters,
$x\in {\cal V}$, $\text{dim}\, {\cal V}\ge 2$.
Most common such parameters  without disturbing  bulk properties
are multiple Aharonov-Bohm fluxes on a genus $g$ Riemann surface. 
When the topological order is non-trivial, there can be  inevitable 
topological 
degeneracies\cite{Wen89}, such as a $q^g$-fold degeneracy of the
FQH state with a filling factor $1/q$ 
on a torus\cite{Yoshioka83,Haldane85p,Haldane85m}.
The degeneracy of a generic ground state 
will be discussed later. Here we just point out that  
one  has to consider non-Abelian gauge structures
arising from it\cite{Wilczek84, Hatsugai04}. 
This is crucial for a numerical concrete characterization of the
topological insulators.
Especially an explicit gauge fixing for
the degenerate multiplet is required
  to perform calculations.\cite{Hatsugai04}.

{\bf  Quantum Spin Systems as  Topological Insulators}\ : 
To describe a characterization of the topological order, 
let us consider a generic translational 
invariant spin-$1/2$ hamiltonian on  a  $d$-dimension 
orthogonal lattice
$
H^P =
 H(h_\ell) =
\sum_{\mb{m}}T^{\mb{m} } 
\, h_\ell((\mb{S}(\mb{r}  _1),\mb{S}(\mb{r} _2),\cdots  )\, 
{T^{\mb{m} } } ^\dagger 
$
 where
$
T^{\mb{m}} = T_{1}^{m_1}\cdots T_{d}^{m_d},\quad
\mb{m} =(m_1,\cdots,m_d)
$
and
${^t\!} \mb{S}(\mb{r})=
{^t\!}  (
{S^x(\mb{r})},
{S^y(\mb{r})},
{S^z(\mb{r})})$
is  a spin-$1/2$  operator at a lattice site $\mb{r} $ and 
$h_\ell$ is a local hamiltonian which depends on several spins 
at $\mb{r}  _1,\mb{r} _2,\cdots $.
It generically breaks several symmetries explicitly such as a parity,
a chiral symmetry, 
and a time reversal symmetry.
The operator $T_{\mu}$ is a translation  in $\mu$-direction,
$
T_{\mu} 
\mb{S}(\mb{r})
T_{\mu} ^\dagger  =\mb{S}(\mb{r}+\mb{a} _\mu)
$ ($\mb{a}_\mu$ is a unit translation in $\mu$ direction).
We use a periodic boundary condition  $T_{\mu}^{L_\mu}=1$
($m_\mu=0,\cdots, d$)
 to avoid disturbing bulk properties by possible edge states.
{\em 
We propose to use twisted boundary conditions for the spin model
and take the twists as the parameters $x$} as discussed below.

{\bf Local Gauge Transformation and Twists} : 
Let us consider a local gauge transformation of a string type,
that is,  local spin rotations 
at a unit cell label $\mb{m} $
as 
$
\mb{S}'_\theta (\mb{r}^{\mb{m} }_\eta) =
\mb{Q}(\gamma )
 \mb{S}(\mb{r}^{\mb{m} }_\eta)
$ with $3\times 3$ matrix 
$
\mb{Q}(\gamma )= e^{\gamma\mb{X}}
$,
$
\gamma={\mb{m}}\cdot \mb{\theta}
$
where 
$
X^{\alpha \beta } = \frac {1}{2}
 i  n^\gamma  {\rm Tr \,} \sigma ^\alpha \sigma ^\beta \sigma ^\gamma 
$,
$\mb{\theta}=(\theta_1,\cdots,\theta_d) $
 and
$\mb{n}=(n_x,n_y,n_z)$ $( |\mb{n} |=1) $ 
 is a fixed rotation axis.
Also  $\eta$ is a label to distinguish intra unit cell spins.
The simplest example is  given by taking $\mb{n}=(0,0,1) $ as 
$
\mb{Q}^z(\gamma ) = 
\left(
\begin{array}{ccc}
\cos \gamma & - \sin \gamma & 0 \\
 \sin \gamma & \cos \gamma  & 0 \\
0 & 0 & 1
\end{array}
\right)$
with 
$\gamma=\mb{m}\cdot \mb{\theta} $.
We further assume the local hamiltonian $h_\ell$ 
is of the gauge interaction type as
\begin{alignat*}{1} 
h_\ell (\mb{S}(\mb{r}_1),\mb{S}(\mb{r}_2),\cdots ) 
=& 
h^G (\mb{\theta}=\mb{0} ;  \{\mb{r}_i-\mb{r}_j\};
\mb{S}(\mb{r}_1 ) ,\mb{S}(\mb{r}_2 ) ,\cdots
)
\\
=& 
h^G (\mb{\theta};\{\mb{r}_i-\mb{r}_j\};
\mb{S}'_\theta(\mb{r}_1 ) ,\mb{S}'_\theta(\mb{r}_2 ) ,\cdots
)
\\
\equiv&
  h^\theta_\ell (\mb{S}'_\theta(\mb{r}_1),\mb{S}_\theta'(\mb{r}_2),\cdots ) 
\end{alignat*} 
with some function $h^G$.
That is, 
{\em the twisting parameters only affect the hamiltonian through 
the relative positions of the local spins. }
Examples of such interactions for the above rotation around $z$-axis are 
$
h_\ell^{\rm pair}=
{^t\!}{\mb{S}}(\mb{r}_1 )
\mb{J} 
\mb{{S}}(\mb{r}_2 )
$
with 
$
\mb{J}=J{\rm diag\, }(1,1,\lambda ) 
$ 
and 
$
 h_\ell^{\rm sb }=
J_c\mb{S} (\mb{r}_1 )\cdot
\big(
\mb{S} (\mb{r}_2 )\times
\mb{S} (\mb{r}_3 )\big)
=J_c \epsilon_{ijk}  \mb{S}_i (\mb{r}_1 )
\mb{S}_j (\mb{r}_2 )
\mb{S}_k (\mb{r}_3 )
$\cite{Wen89wwz}.
They transform respectively as 
$
h^{\rm pair} = \frac {J}{2} 
(e^{-i(\theta_1-\theta_2) }{S_1}_{\theta_1}^+{S_2}_{\theta_2}^- +
h.c.)
+ 
\lambda {S_1}_{\theta_1}^z{S_2}_{\theta_2}^z
$,
$
h^{\rm sb} = \frac {J_c}{2} {S_1}_{\theta_1}^z (i
e^{-i(\theta_2-\theta_3) }{S_2}_{\theta_2}^+{S_3}_{\theta_3}^- + h.c.)
+(\text{cyclic perm.})
$,
$\theta_i=\mb{m}_i\cdot \mb{\theta} $, $i=1,2,3$ 
where $\mb{m}_i $ is a unit cell labeling of the spin $\mb{S}(\mb{r} _i) $.

The hamiltonian $H^P$ is periodic in the original
  representation by $\mb{S} $'s but is
not periodic in the one by twisted $\mb{S}_\theta $'s
as
 $\mb{S}'(T^{L_\mu} \mb{r} )={ T'}_\mu^{L_\mu}\mb{S}'(\mb{r} ) $ with
$
{T'}_\mu^{L_\mu} =
 \exp(- \hat n_\mu \theta_\mu L_\mu \mb{X}  )
$.

Now let us define a translational invariant
twisted hamiltonians $H^T$ by a representation by
the twisted $\mb{S}'_\theta$ 
as
$
H^T(\mb{\theta}) =
 \sum_{\mb{m} } {T'}^{\mb{m} }
 h^\theta_\ell
 (\mb{S}'_\theta(\mb{r}_1),\mb{S}_\theta'(\mb{r}_2),\cdots ) 
{{T'}^{\mb{m}} }^\dagger 
$
 with a periodic boundary condition
$
{T'}_\mu^{L_\mu} = 1
$.
In the original spin operators $\mb{S} $, $H^T$ is given by $H^P$ 
with the twisted  boundary condition
$T_\mu^{L_\mu} = \exp(+  \theta_\mu L_\mu  \mb{X}  )$.

The two hamiltonians,
  $H^P$ and $H^T$,
are both translationally invariant 
in   representations by
 $\mb{S} $ and $\mb{S}'_\theta $ respectively.
One may expect an  macroscopic ${\cal O}(V)$ energy difference between 
their
ground state energies.
 However, as discussed, 
the contribution should be at most 
${\cal O}(|\partial_\mu V|)$ 
due to the gauge invariance where $|\partial_\mu V|$ is a  (hyper) area
of the system perpendicular to the $r_\mu$-axis
where $V=L_1\cdots L_d$. 
That is, the difference of the energy should be a finite size effect.
Thus the difference between 
  $H^P$ and $H^T$ is negligible in the thermodynamic limit $V\to
\infty$ when we discuss the bulk properties in a usual manner.

Another important point for the present  construction is that 
the twisted hamiltonian $H^T$ is translational invariant 
in the $\mb{S}_\theta $ representation.
Then edge states never appear
in any representation,
which is especially 
important to pick up only bulk properties through probes by 
twisted boundary conditions.

\begin{figure}
\begin{center}
\includegraphics[width=5.0cm]{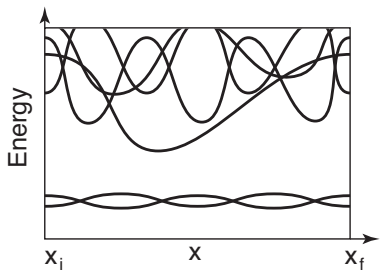}%
\end{center}
 \caption{Schematic spectral flow with parameters $x$.
\label{multiplet}}
 \end{figure}

{\bf Degeneracies and Ground State Multiplet}\ :
A topological order on a non-zero genus Riemann surface
 is one of the 
 reasons for the ground state degeneracy, which is the topological
 degeneracy\cite{Wen89}. 
The simplest example is just  a manybody state with two-dimensional 
periodic boundary conditions\cite{Yoshioka83,Haldane85p}.
Also if the system has a standard symmetry breaking, such as 
ising orders,
a finite system has almost degenerate ground states corresponding to
 linear combinations of the symmetry broken states\cite{Hatsugai97a}.
For a finite system, 
the degeneracy can be lifted and the splitting is estimated as
$\approx e^{-C V}$
for the symmetry broken states.
Also if the ground state has a finite spin moment which may not be
macro as a ferromagnet, there occurs a spin degeneracy.
Some of these degeneracies can be approximate for a finite system
and may be sensitive to the boundary condition and twisting
parameters, such as  $\theta_\mu$'s.
In these cases, the lowest energy gap is not a physical one and
may vanish in the thermodynamic limit.
The physical energy gap for the bulk is an energy gap above
 these almost degenerate states.
We define a ground state multiplet
by a collection of  these almost degenerate states near the ground state
and define a Chern number for this ground state multiplet. 
(See the Fig.\ref{multiplet})
\footnote{ 
Since the hamiltonian 
$H^T(\mb{\theta}  )$ is translationally invariant,
${[}H^T,T'_\mu{]}=0$. That is $T'_\mu = e^{i k_\mu}$ with 
$k_\mu=  {2\pi m_\mu}/{L_\mu}$, $m_\mu=0,\pm 1,\pm 2,\cdots$,
the  ground state multiplet may compose of states with
different $k_\mu$. 
We may define the Chern numbers using these conserved quantum numbers.
However it is natural to discuss (almost) 
degenerate states as a ground state multiplet\cite{Hatsugai04}.
}
Since the two hamiltonians $H^P$ and $H^T(\mb{\theta})$ 
differ only boundary terms, bulk properties of the two should be the
same. 
Then, for the topological insulators,
the energy gap above the ground state multiplet is 
stable against  perturbations.
If the ground state multiplet
 is well separated from the above in a finite system,
{\em
 we do not need to take the thermodynamic limit. 
}

The twist we proposing is a boundary perturbation in a particular
representation. However it also preserve the translational symmetry up to a 
unitary equivalence.
Then, based on a discussion of  edge state picture, 
we expect an energy separation of the topological degeneracy,
 that is, a band width of the multiplet by the twist as 
$e^{-L/\xi}$ where $L$ is a minimum linear dimension of the system 
and $\xi$ 
is a typical length scale of the ground state multiplet.
It can be different from the conventional broken symmetric cases.

{\bf Chern Numbers for the Spins }\ : 
Let us define a total parameter space by
${\cal V}=
\{(\theta_1,\cdots,\theta_d)|\theta_\mu\in [0,2\pi/L_\mu]\}$.
Since
$
\exp( 2\pi  \mb{X}  )= \mb{I}_3 $,
we have
$H^T(\mb{\theta}+(\cdots,0,\theta_\mu+2\pi/L_\mu,0,\cdots )
=H^T(\mb{\theta})
$ in the $\mb{S} $-representation.
Then the twisted hamiltonian $H^T(\mb{\theta} )$
is well defined 
on 
$\cal V$ without boundaries as ${\cal V} =T^d$.
\footnote{ 
When we use a representation by $\mb{S}_\theta $,
$H^T(\mb{\theta}+(\cdots,0,\theta_\mu+2\pi/L_\mu,0,\cdots )
\neq
H^T(\mb{\theta})
$ and it 
is not well defined  on the surface $\cal S$.
It is just well defined up to an
unitary equivalence.}
Any two dimensional integration surface ${\cal S}(\subset {\cal V})$ 
without boundaries is used to define the Chern numbers.
${\cal S}=T_{ij}^2=\{(\theta_i,\theta_j)\}$ is the simplest example.
Now define a ground state multiplet $\mb{\Psi}(x),x\in {\cal V}$.
It is a $N\times q$ matrix as
$\mb{\Psi}(x) =
 (\psi_1(x),\cdots,\psi_q(x))
$
 with
$
H^T(x)\psi_j(x)=\epsilon_{j} \psi_j(x), j=1,2,\cdots$,
 $\epsilon_{i}\le \epsilon_{j},(i<j) $,
where $\psi_j$ is a column vector in a many spin Hilbert space with a
dimension $N$
and $q$ is a dimension of the ground state multiplet.
The  {\em generic energy gap condition} for the multiplet is given as
$
\epsilon_{q}(x)< \epsilon_{q+1}  (x),\forall x\in {\cal S}
$. 
This is a definition of the topological insulators.

Define a non-Abelian connection one-form $\mb{\cal A} $ 
which is an $q\times q$ matrix as 
$
\mb{\cal A} = \mb{\Psi} ^\dagger d \mb{\Psi} 
$ 
and a field strength two-form
$
{\mb{\cal F}} = d {\mb{\cal A}} + {\mb{\cal A}} ^2
$.
The first Chern number\cite{eguchi80} is then defined by
$
C_{\cal S} =
  \frac {1}{2\pi i} \int_{\cal S} {\rm Tr \,} {\mb{\cal F}} 
= \frac {1}{2\pi i} \int_{\cal S} {\rm Tr \,} d {\mb{\cal A}} 
$. 
It is a topological integer which is stable against perturbation
unless the generic gap collapses.
We use these integers depending on  a choice of ${\cal S}$ to 
characterize the topological orders.
Changing a basis within the multiplet space, 
$ 
\mb{\Psi}'(x) =  \mb{\Psi} (x) \mb{\omega} (x)
$,
($\mb{\omega} \mb{\omega} ^\dagger = \mb{I}_q  $)
gives a gauge transformation 
$
\mb{\cal A } ' = \mb{\Psi} ' d \mb{\Psi} ' 
= 
\mb{\omega } ^{-1}  \mb{\cal A} \mb{\omega }  + \mb{\omega } ^{-1} 
d \mb{\omega }
$ and
$
{\mb{\cal F}} ' =
\mb{\omega } ^{-1}
{\mb{\cal F}} 
\mb{\omega } 
$
\cite{Wilczek84,Hatsugai04}.
 The Chern number is a gauge invariant but
we need to fix the gauge to evaluate this expression\cite{Hatsugai04}.
Take a generic arbitrary multiplet, $\mb{\Phi}$, and 
define an overlap matrix as
$
{\mb{ O}}_\Phi = \mb{\Phi} ^\dagger \mb{P} \mb{\Phi}
$
where $\mb{P}=\mb{\Psi} \mb{\Psi} ^\dagger    $ is
a projection into the ground state multiplet 
which is a gauge invariant. 
Then define regions
${\cal S}_R^\Phi, R=1,2,\cdots$,  as  (infinitesimally) small
neighborhoods  of 
zeros  of  $\det \mb{ O}_\Phi(x) $ 
and  ${\cal S}^\Phi_0$ as a rest of ${\cal S}$.
Then the first Chern number
is written as
$
C_{\cal S} =
 - N_\Omega^T({\cal S}) 
=-\sum_{R\ge 1} n_\Omega^R({\cal S}_R^\Phi)
$,
$
 n_\Omega^R({\cal S}_R^\Phi) =   \frac {1}{2\pi} \oint _{\partial
   {\cal S}^\Phi_R}
d'  \Omega
$. 
The  field $\Omega $ is defined as 
$ \Omega (\tilde {\mb{\Phi}} ,{\mb{\Phi}} ) = 
 {\rm Arg\,} 
\det \mb{\tilde\Phi}^\dagger  \mb{P}\mb{\Phi}
=  \text{Arg}\det {\mb{\eta} } -\text{Arg}\det \tilde{\mb{\eta} } $
where 
$\tilde{\mb{\Phi}} $ is also another generic arbitrary multiplet,
$\mb{\eta}  =  \mb{\Psi} ^\dagger \mb{\Phi}$ and
$\tilde{\mb{\eta}}  = \mb{\Psi} ^\dagger\tilde{\mb{\Phi}}$.
The matrices $\mb{\eta} $ and $\tilde{\mb{\eta}} $  depend on the
choice of the multiplet $\mb{\Psi} $ but
the difference of the arguments is a gauge 
invariant.

The field $\Omega$ 
depends on  a
 choice of $\mb{\Phi} $ and
 $\tilde{\mb{\Phi}} $  but
the total vorticity $N_\Omega^T({\cal S})$
  is a gauge invariant and independent of the choice.
{\em 
The field $\Omega$ reflects 
a phase sensitivity of the multiplet 
by the twist}
when one fixes $\mb{\Phi} $ and  $\tilde{\mb{\Phi}} $.
It is illustrative to show $\Omega $ and it supplies
 information of the ground state multiplet. 
Also  when the integration surface ${\cal S}$ is contractible to a point
keeping a generic energy gap, the Chern number vanishes from a 
topological stability.

\begin{figure}
\includegraphics[width=8.0cm,clip]{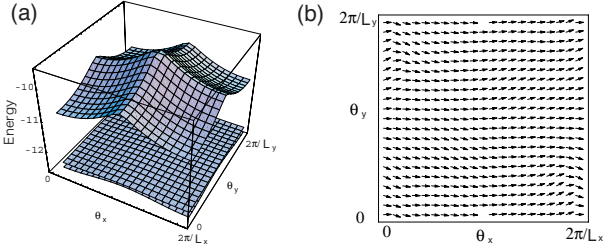}%
 \caption{ 
(a)Three lowest energies of the XXZ model on a $4\times 4$ 
square lattice with twists in the total $S_z=0$ sector is shown.
 ($\lambda=1.3$)
(b)
A  field $\Omega $ of the ground state multiplet composed of  
the lowest two eigen states 
for some choice of $\mb{\Phi},\tilde{\mb{\Phi}}$.
\label{ising}}
 \end{figure}

{\bf Ex.1:Two-Dimensional Spin Model}\ :
The present formulation can be effective 
for characterization of topological ordered phases
in any dimensions,
such as chiral spin states\cite{Wen89wwz}.
{\em 
To have a finite Chern number, one needs to break time reversal symmetry
}
 as for the quantum Hall states
\footnote{ 
Non-trivial examples with finite Chern numbers
for spin models are under studies in collaboration with
X.-G. Wen. ( unpublished )
}.
The simplest example of $h_\ell$ 
can be a sum of local pair-spin interactions $ h_\ell^{\rm pair}$
with  a symmetry breaking term $ h_\ell^{\rm sb}$
discussed above as 
$
h_\ell= \sum_{\rm pair}h_\ell^{\rm pair}+h_\ell^{\rm sb}
$.
Here  let us just show  an example
with a degeneracy to show the present general procedure. 
Considering only a nearest neighbor exchange interaction
and assume $\mb{n}=(0,0,1) $,
the local hamiltonian is given as
$
h_\ell^{{\theta}} =
\sum_{\mu=x,y}
J  \big(
\frac {1}{2}  (
e^{-i \theta^\mu  }
{S_\theta}_+^\mu 
{S_\theta}_-
+
e^{i \theta^\mu  }
{S_\theta}_-^\mu 
{S_\theta}_+)
+ \lambda {S_\theta}_z^\mu  {S_\theta}_z
\big)
$ 
where
 $\mb{S}_\theta^\mu= \mb{S}_\theta( T^{\mu}\mb{r})$
and 
 $\mb{S}_\theta= \mb{S}_\theta( \mb{r})$.
In the case, the twisted boundary condition for $H^T(\mb{\theta} ) $
in $\mb{S} $ is  given by the following matrix
$
T_\mu^{L_\mu}  = \mb{Q} ^z(\gamma)
$
with
$\gamma= \theta_x L _x+\theta_y L _y$.
Then we can take a 2 dimensional torus $T^2=\{(\theta_x,\theta_y)|0\le
\theta_\mu\le 2\pi/L_\mu\}$ for the integration surface ${\cal S}$.
This is  a
nearest neighbor $XXZ$ model 
on a square lattice 
with twists.
With an ising anisotropy, $\lambda>1$,
the ground state of an infinite system has 
a long range order and
has a finite energy gap.
We show numerical results for a system with $J=1$ and $\lambda=1.3$.
The  ground state of a finite size system is given by
a bonding state between two symmetry broken states with
antiferromagnetic (ising)
order. The next lowest state is an anti-bonding state of them 
and the energy separation between is
expected to be $\propto e^{- { L_xL_y}/{\xi^2}}$  where $\xi$ 
is a typical length scale. 
A physical 
 ising gap to flip one ordered spin is given the one above,
that is, the second lowest one.
Therefore the ground state multiplet is composed of the two
low lying states including the finite size ground state.
(See Fig.\ref{ising}(a))
The field $\Omega  $ of the two dimensional ground state
multiplet 
is shown in Fig.\ref{ising}(b).
As discussed in the reference\cite{Hatsugai04}, 
the Chern number is a sum of the vorticity at the zeros of 
$\det {\mb{O}}_\Phi=|\det \mb{\eta}|^2 $. 
In the present example, it is $0$ as expected for a chiral symmetric system.

{\bf Ex.2, Manyparticle States in the  First Quantized Form }\ : 
The same procedure is also applied for a manyparticle state in the
first quantized form,
such as the generic FQH States 
$
\Psi_k(x;
\mb{r}_1,\mb{\sigma }_1;
\cdots;
\mb{r}_N,\mb{\sigma }_N
  )
$
where $k$ denotes a label of the (topological) degeneracy
of the ground state multiplet and
 $x=(\theta_x,\theta_y)$ is a set of parameters specifying
 twisted boundary conditions 
on a torus\cite{Haldane85p}.
By taking a reference multiplet as 
$
\Phi_\xi
(
\mb{r}_1,\mb{\sigma }_1;
\cdots;
\mb{r}_N,\mb{\sigma }_N
)
=
\delta_{\sigma_1 \sigma_1^\xi}
\delta(\mb{r}_1 -\mb{r}^\xi_1)
\cdots
\delta_{\sigma_N \sigma_N^\xi}
\delta(\mb{r}_N -\mb{r}^\xi_N),
\xi=1,\cdots, q
$. 
The Chern number for the degenerate multiplet 
is given by
the  field 
$
\Omega (x)= {\rm Arg} \det {\tilde{\mb{\eta}} } ^\dagger {\mb{\eta}}
$
with
$
\{\mb{\eta}\}_{\xi k}  = 
\Psi_k(x;
\mb{r}_1^\xi,\mb{\sigma }_1^\xi;
\cdots;
\mb{r}_N^\xi,\mb{\sigma }_N^\xi
  )
$
and 
$
\{ \tilde{\mb{\eta}}\}_{\tilde\xi k}  = 
\Psi_k(x;
\mb{r}_1^{\tilde\xi},\mb{\sigma }_1^{\tilde\xi};
\cdots;
\mb{r}_N^{\tilde\xi},\mb{\sigma }_N^{\tilde\xi}
  )
$
where $k,\xi,\tilde \xi$ run over $\{1,\cdots,q\}$.
The Chern number is evaluated as 
a total vorticity of $\Omega$ at the zeros of 
$|\det \mb{\eta}|^2 $\cite{Hatsugai04}.

We thank X.-G. Wen for very fruitful discussion.
Part of the  present work  was supported by a Grant-in-Aid from the
Japanese Ministry of Education, Culture, Sport, Science and Technology, and 
the JFE 21st Century Foundation. 
\vfill

\end{document}